\documentclass[doublespacing]{elsart}

\usepackage{epsfig}

\usepackage{amssymb}

\begin{document}

\begin{frontmatter}




\title{AN EMPIRICAL ANALYSIS OF MEDIUM-TERM INTEREST RATES}


\author[label1]{Tiziana Di Matteo},
\author[label2,cor1]{Enrico Scalas},
\author[label3]{Marco Airoldi}
\address[label1]{INFM - Dipartimento di Fisica "E. R. Caianiello",
Universit\`a degli Studi di Salerno, I-84081 Baronissi (Salerno),
Italy.}
\address[label2]{Dipartimento di Scienze e Tecnologie Avanzate,
Universit\`a del Piemonte Orientale, I-15100 Alessandria, Italy, \\ phone:
+39 0131  283854, fax: +39 0131 254410.}
\address[label3]{Risk Analysis $\&$ Integration Unit, Risk Management $\&$
Research,
Banca Intesa, Milano, Italy.}

\begin{abstract}
In the present paper, an empirical study of LIBOR (London
Interbank Offered Rate) data is presented. In particular, a data
set of interest rates from $1997$ to $1999$, for two different
currencies and various maturities, is analyzed. It turns out that
the random behavior of the daily increments for the interest rates
series is non-Gaussian and follows a leptokurtic distribution.
\end{abstract}

\end{frontmatter}

\newpage

\section{Introduction}  \label{s.1}

In financial theory and practice, interest rates are a very
important subject which can be approached from several different
perspectives.

The classical theoretical approach models the term structure of
interest rates using stochastic processes and the no arbitrage
argument. Various models have been proposed and can be found in Pagan
(1997), Bouchaud (1997), Rebonato (1998), Wilmott (1999), Hull (2000),
Mantegna (2000), Dacorogna (2001), Vasicek (1977), Cox (1985), Heath
(1992), Hull (1993)
\cite{Pagan,BouchaudPot,Rebonato,Wilmott,Hull,LibrMant,LibDac,Vasicek,Cox,Heath,HullWhite}.
Although they provide analytical formulas for the pricing of
interest rate derivatives, the implied deformations of the term
structure have a Brownian motion component and are often rejected
by empirical data (see Chan (1992) \cite{Chan}). The inadequacies of the
Gaussian
model for the description of financial time series has been
reported since a long time ago by Mandelbrot (1963), see also Mandelbrot
(1997) \cite{Mandelbrot,MandelbrotB}, but thanks
to the availability of large sets of financial data, the interest
on this point has risen recently in Mantegna (2000), Dacorogna (2001),
Sornette (2000), Muller (1995), Mantegna (1994), Ghashghaie (1996),
Mantegna (1997)
\cite{LibrMant,LibDac,SornSim,Muller,Mantegna,Ghashghaie,MantegnaSta}.
In particular, the fat-tail property of the empirical distribution
of price changes has been widely documented and is a crucial
feature for monitoring the extreme risks (Embrechts (1997), Embrechts
(1999), McNeil (1998), McNeil (1999)
\cite{Embrechts1,Embrechts2,McNeil1,McNeil2}). Many of the recent
studies concern high frequency data of stock indexes or exchange
rates (Muller (1995), Ghashghaie (1996), Mantegna (1997), Muller (1996)
\cite{Muller,Ghashghaie,MantegnaSta,MullerDac}).

Also for interest rates, up to now, no universally accepted theory
has been obtained for the description of experimental data.
Dacorogna has investigated the intraday behavior of the interest
rate markets (Muller (1996), Piccinato (1997), Ballocchi (1997), Ballocchi
(1999) \cite{MullerDac,Piccinato,Balocchi,Baloc}). Bouchaud
et al. (Bouchaud (1997), Matacz (1999), Matacz (2000)
\cite{Bouchaud,Matactz1,Matactz2}) have considered the
interest rate curve as a vibrating string subject to random shocks
along its profile. Along these lines, Sornette (1998) \cite{Sornette},
has derived a general condition that the partial differential
equations governing the motion of such string must obey in order
to account for the condition of absence of arbitrage. Jean Nuyts
(1999) \cite{Nuyts} has developed a phenomenological approach based on
Pad\'e Approximants. In the work of Jamshidian (1997) \cite{Jamshidian},
a self-contained theory is presented for pricing and hedging LIBOR
and swap derivatives by arbitrage. Appropriate payoff homogeneity
and measurability conditions are identified which guarantee that a
given payoff can be attained by a self-financing trading strategy.
An important recent development in the pricing of interest rate
derivatives is the emergence of models that incorporate lognormal
volatilities for forward LIBOR or forward swap rates while keeping
interest rates stable. Zhao et al. (1999) \cite{Zhao} have introduced
methods for discretizing these models giving particular attention
to precluding arbitrage among bonds and to keeping interest rates
positive even after discretization.

In this framework, we have empirically studied the probability
density distribution of LIBOR, in order to characterize the
stochastic behavior of the daily fluctuations.

In Section \ref{s.2}, we present the data set. The data are then
analyzed and the results are given in Section \ref{s.3}. Section
\ref{s.4} contains a discussion and some final remarks.

\section{The data set} \label{s.2}

LIBOR stands for the London Interbank Offered Rate and is the rate
of interest at which banks are willing to offer deposits to other
prime banks, in marketable size, in the London interbank market.

BBA (British Bankers' Association) LIBOR \cite{bba} is the most
widely used benchmark or reference rate. It is defined as the
interest rate at which deposits are perceived to be generally
available in the London interbank at $11AM$ (Greenwich time).

Moreover, it is used as the basis for settlement of interest rate
contracts on many of the world's major future and option exchanges
as well as most Over the Counter (OTC) and lending transactions.

The BBA, advised by senior market practitioners, maintains a
reference panel of $16$ banks. The purpose is to produce a
reference panel of banks reflecting the balance of the market by
country and by type of institution. Individual banks are selected
within this principle on the basis of reputation, scale of market
activity and perceived expertise in the currency concerned. The
BBA surveys the panel's market activity and publishes their market
quotes on-screen. The highest four and lowest four market quotes
are disregarded and the remaining eight quotes are averaged: the
resulting spot fixing is the BBA LIBOR rate. The quotes from all
$16$ panel banks are published on-screen to ensure transparency.
BBA LIBOR is compiled each working day by an electronic vendor
(currently Dow Jones Telerate) and broadcast through ten
international distribution networks including Reuters, Dow-Jones
Telerate, Knight-Ridder, and Bloomberg.

BBA LIBOR fixings are provided in seven international currencies:
Pound Sterling, US Dollar, Japanese Yen, Swiss Franc, Canadian
Dollar, Australian Dollar, EURO. LIBOR rates are fixed for each
currency at monthly maturities from one month to $12$ months.
Rates shall be contributed in decimal to at least two decimal
places but no more than five.

\begin{figure}
\begin{center}
\mbox{\epsfig{file=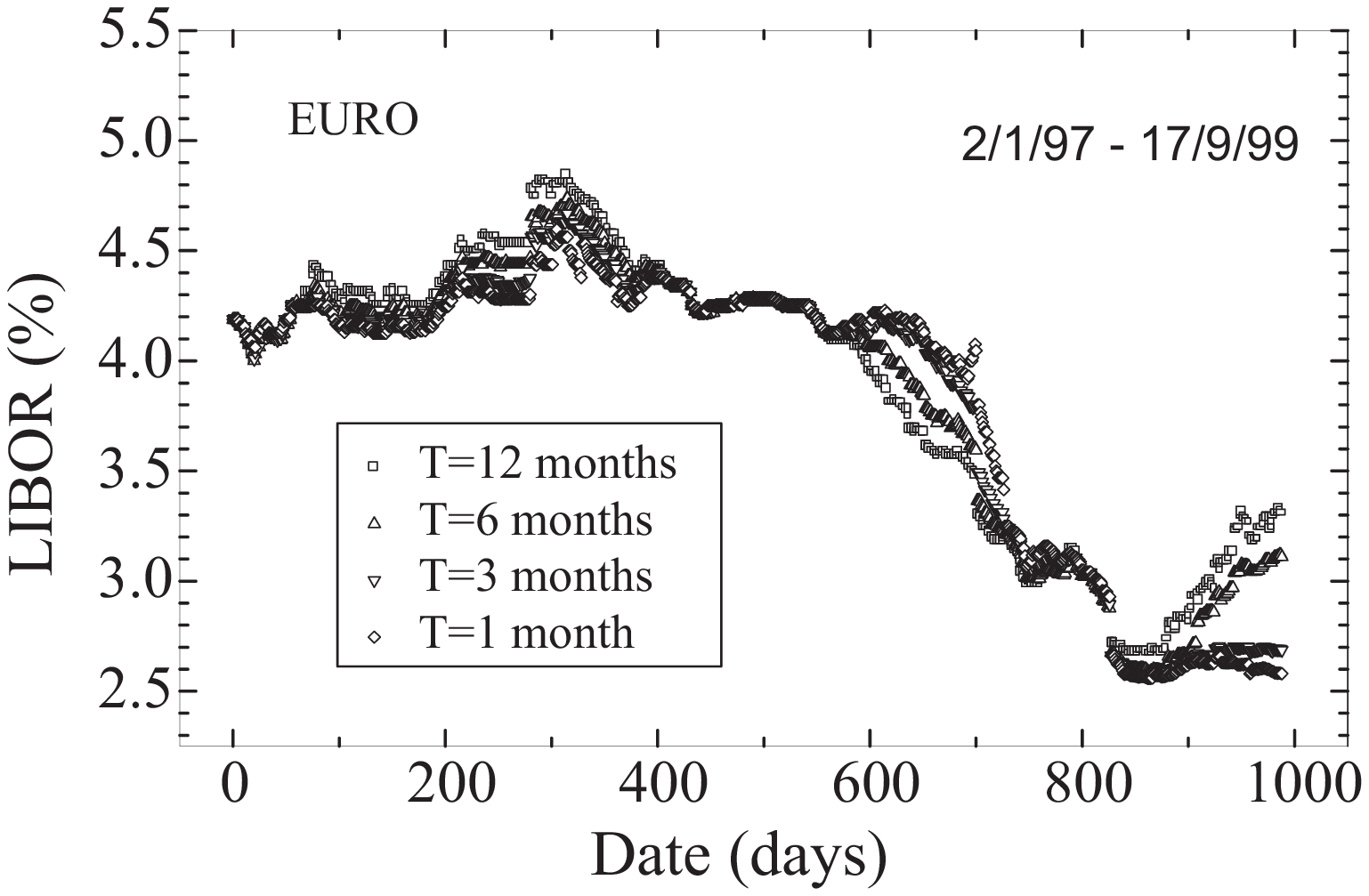,width=5in,angle=0}}
\end{center}
\caption{LIBOR interest rates $r(T,t)$ as a function of the
current date $t$, for the maturities $T$=$1$, $3$, $6$ and $12$
months and EURO currency.}
\label{f.1}
\end{figure}

\begin{figure}
\begin{center}
\mbox{\epsfig{file=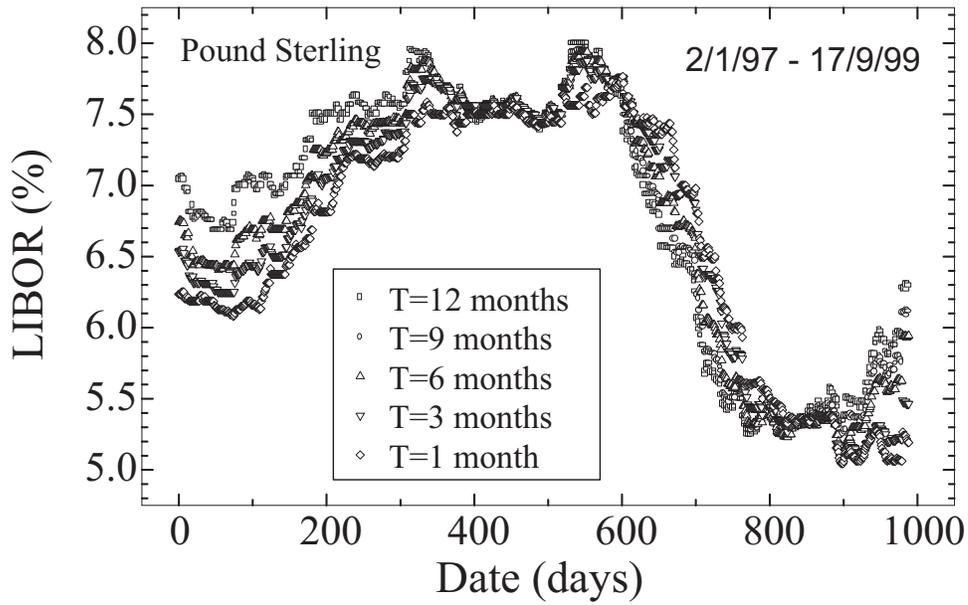,width=5in,angle=0}}
\end{center}
\caption{LIBOR interest rates $r(T,t)$ as a function of the
current date $t$, for the maturities $T$=$1$, $3$, $6$, $9$ and
$12$ months and Pound Sterling currency.}
\label{f.2}
\end{figure}

In the following, we have analyzed a data set of LIBOR interest
rates $r(T,t)$, where $T$ is the maturity date and $t$ the current
date, for EURO and Pound Sterling. These data are shown in
Fig.\ref{f.1} and Fig.\ref{f.2} where $t$ goes from January 2,
$1997$ to September $17$, $1999$, and $T$ assumes the following
values: $1$, $3$, $6$, $9$ and $12$ months for the Pound Sterling
and $1$, $3$, $6$, $12$ months for the EURO.

\begin{figure}
\begin{center}
\mbox{\epsfig{file=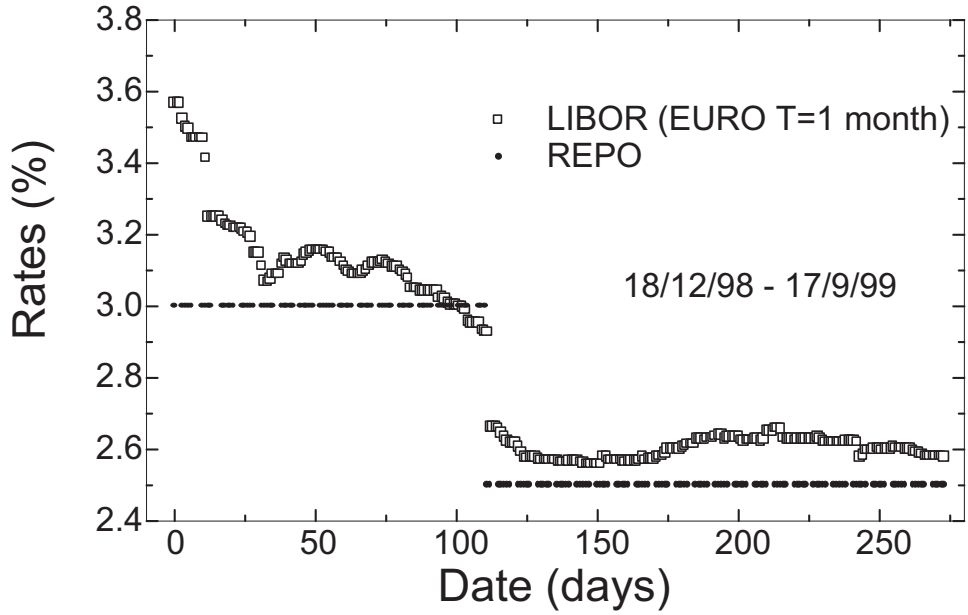,width=5in,angle=0}}
\end{center}
\caption{LIBOR compared to REPO for EURO.}
\label{f.3}
\end{figure}

\begin{figure}
\begin{center}
\mbox{\epsfig{file=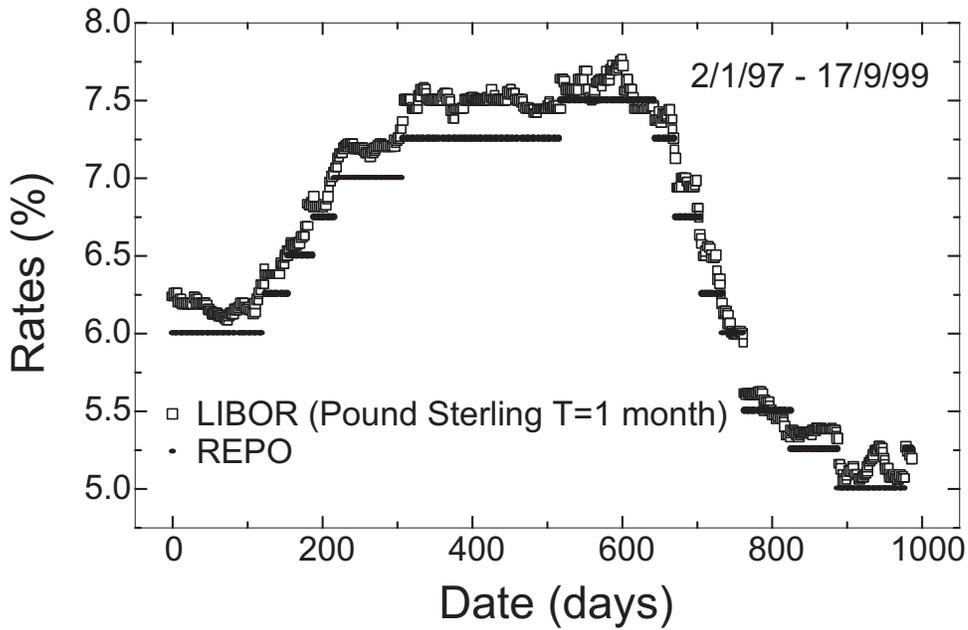,width=5in,angle=0}}
\end{center}
\caption{LIBOR compared to REPO for Sterling Pound.}
\label{f.4}
\end{figure}

\begin{figure}
\begin{center}
\mbox{\epsfig{file=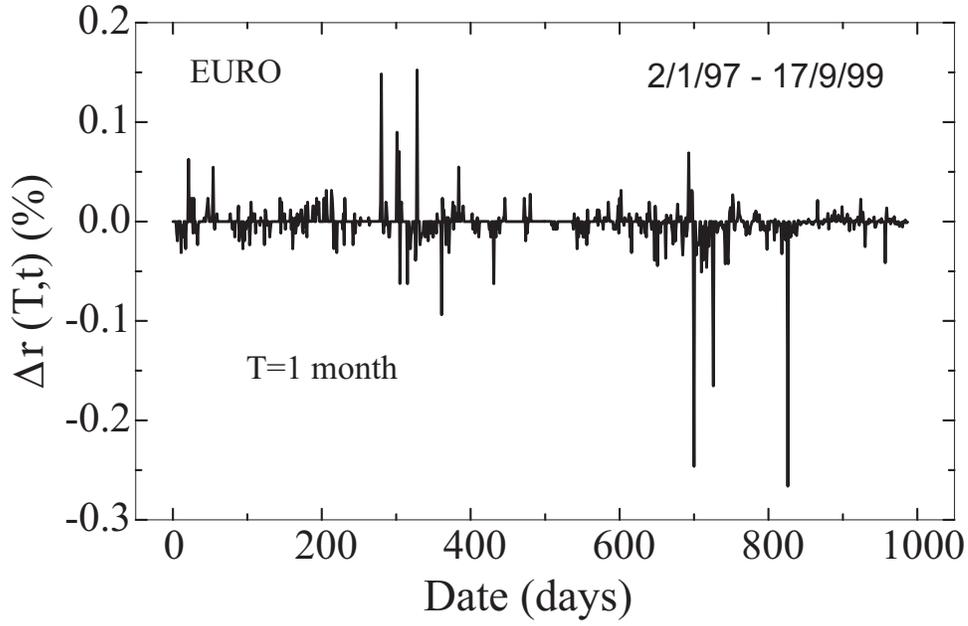,width=5in,angle=0}}
\end{center}
\caption{1-month LIBOR increments as a function of the current
date for EURO.}
\label{f.5}
\end{figure}

\begin{figure}
\begin{center}
\mbox{\epsfig{file=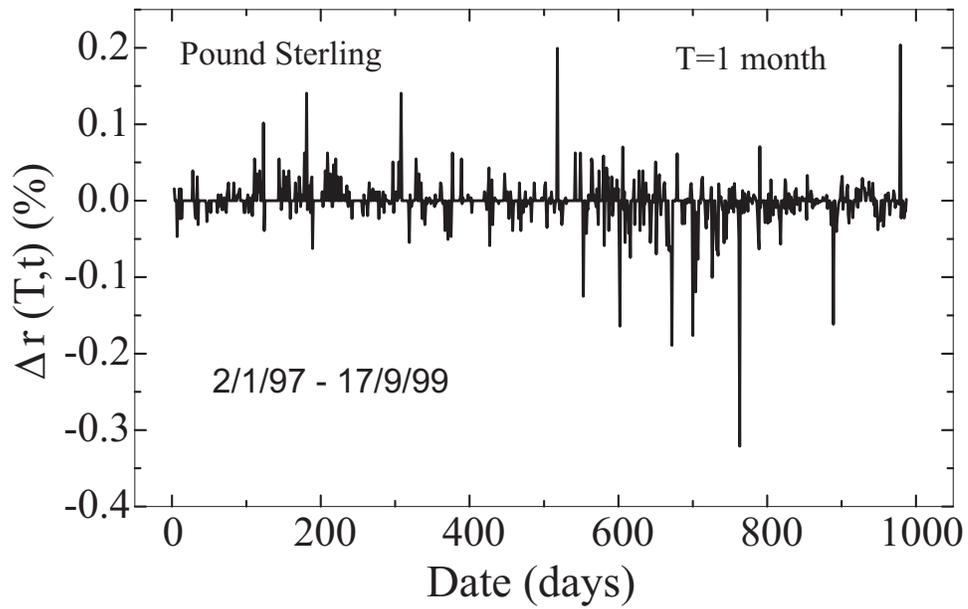,width=5in,angle=0}}
\end{center}
\caption{1-month LIBOR increments as a function of the current
date for Sterling Pound.}
\label{f.6}
\end{figure}

In Figs.\ref{f.3}, \ref{f.4} the 1-month LIBOR is compared to the
interest rates fixed by Central Banks at that time, namely the
REPO (repurchase agreement) data. BBA LIBOR follows the trend
determined by the decisions of Central Banks. In order to roughly
eliminate these trends, in Figs.\ref{f.5} and \ref{f.6} the
interest rates differences $\Delta r(T,t)$=$r(T,t+\Delta
t)-r(T,t)$, with $\Delta t$ being $1$ day and $T$=$1$ month, are
plotted as a function of the current date, for the EURO and the
Pound Sterling, respectively. A similar behavior is also found for the
other maturities. Some large oscillations of $\Delta
r$ are induced by Central Banks. In any case, $\Delta r$ heavily
fluctuates around zero.

\section{Results} \label{s.3}

In this section, we focus the attention on the probability
distribution behavior of the interest rates increments $\Delta
r(T,t)$ defined in the previous section. To this purpose, we
estimate $\Psi (\Delta r)$, the complementary cumulative
distribution function of the daily interest rates increments,
defined as:

\begin{equation}
\label{uno} \Psi (\Delta r)=1-\int_{-\infty}^{\Delta r} p(\eta)
d\eta
\end{equation}

\noindent where $p$ is the probability of $\Delta r(T,t)$.

\begin{figure}
\begin{center}
\mbox{\epsfig{file=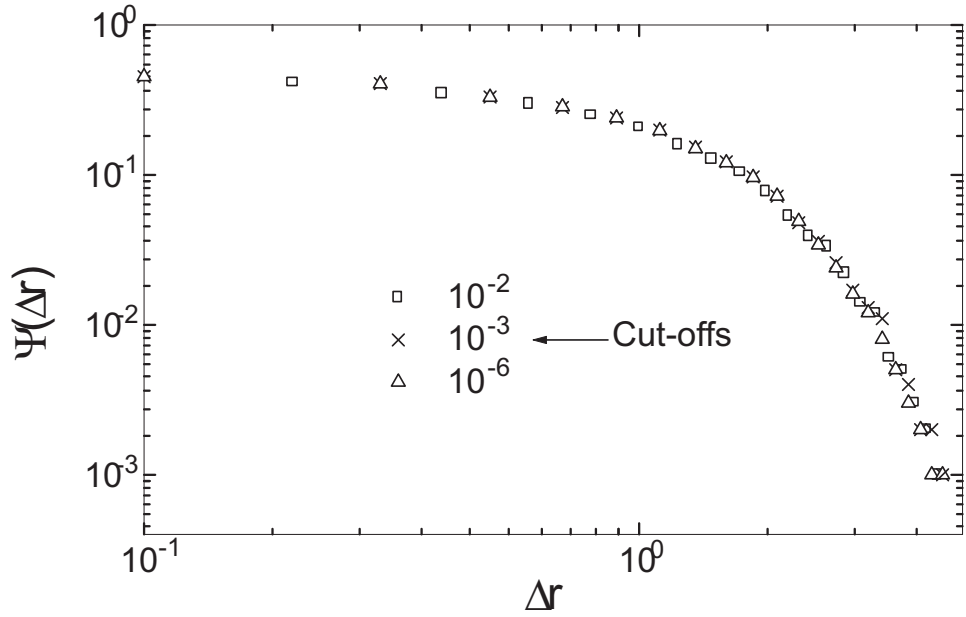,width=5in,angle=0}}
\end{center}
\caption{The complementary cumulative distribution function of a
simulated Gaussian stochastic process using different decimal digit
precisions (cut-offs). }
\label{f.7}
\end{figure}

Because LIBOR data are supplied with only few decimal digits, it
is interesting to examine the effects of different data cut-offs
in the behavior of $\Psi (\Delta r)$. In Fig.\ref{f.7}, we plot
the complementary cumulative distribution function for a simulated
Gaussian
stochastic process using data characterized by three different
decimal digit precisions. It turns out that the numerical rounding
does not influence the results.

\begin{figure}
\begin{center}
\mbox{\epsfig{file=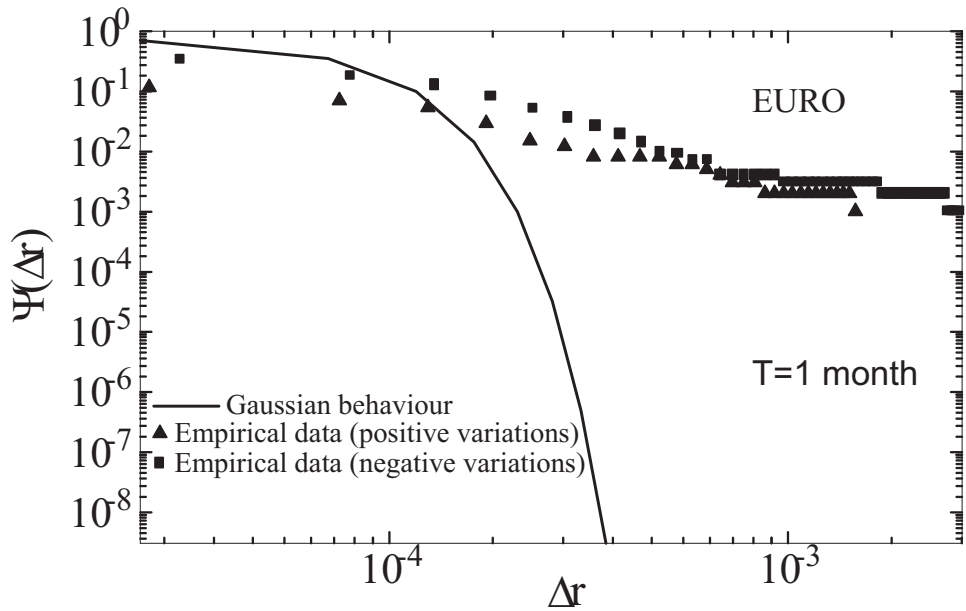,width=5in,angle=0}}
\end{center}
\caption{The complementary cumulative distribution function of
LIBOR increments for EURO. $T$=1 month and $\Delta t$=1 day.}
\label{f.8}
\end{figure}

\begin{figure}
\begin{center}
\mbox{\epsfig{file=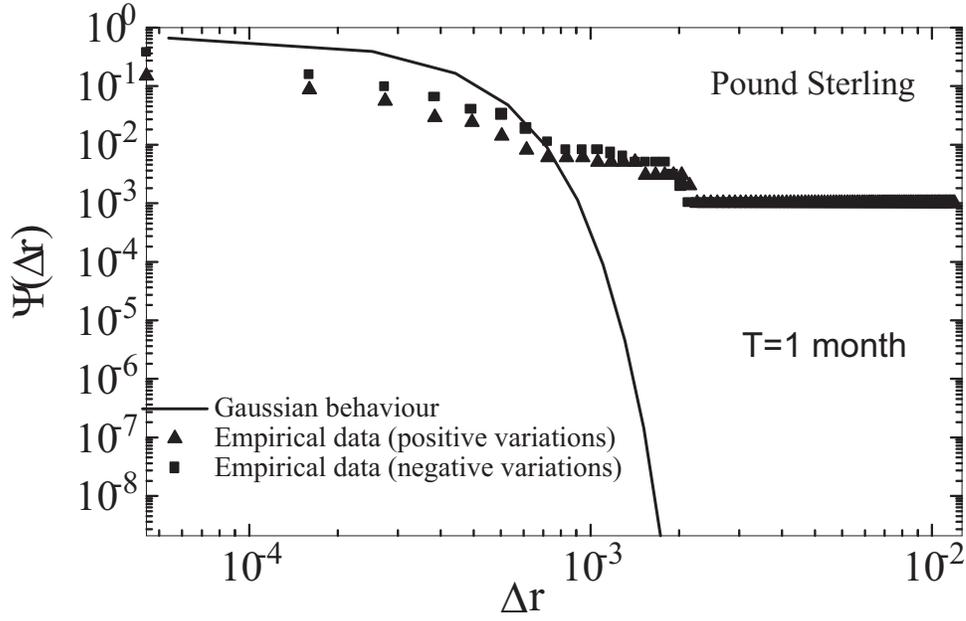,width=5in,angle=0}}
\end{center}
\caption{The complementary cumulative distribution function of
LIBOR increments for Sterling Pound. $T$=1 month and $\Delta t$=1
day.}
\label{f.9}
\end{figure}

Figs.\ref{f.8} and \ref{f.9} show the tail distribution behaviors
in the case of EURO and Sterling Pound, respectively. In
particular, in Fig.\ref{f.8}, we report the empirical results
obtained estimating the probability density function of both
positive and negative LIBOR increments with $\Delta t$=$1$ day and
$T$=$1$ month. These empirical curves are slightly asymmetric and
the negative variations are more probable than the positive one.
In the same figure, these two curves are compared with the
equivalent (i.e. with the same average and standard deviation)
Gaussian complementary cumulative distribution. The non-Gaussian
behavior is also evident from Fig.\ref{f.9}. In both
Figs.\ref{f.8} and \ref{f.9}, the empirical LIBOR data exhibit a
{\it fat tail} or {\it leptokurtic} character, which is present for the
other maturities as well. These observations
indicate that the random behavior of $\Delta r(T,t)$ is
non-Gaussian and that using a Gaussian probability density
function leads to underestimating the probability of large
fluctuations.

For a better understanding of what kind of stochastic process we
are dealing with, we present some results on the power spectral
density behavior (Kay (1981) \cite{Kay}).

\begin{figure}
\begin{center}
\mbox{\epsfig{file=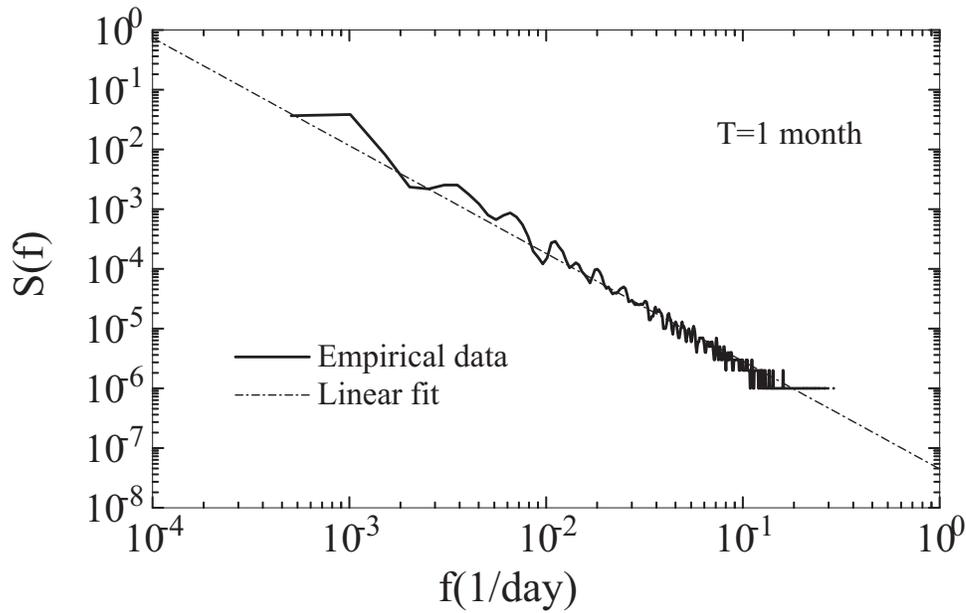,width=5in,angle=0}}
\end{center}
\caption{Power spectrum of $r(T,t)$ for the EURO with $T=1$
month.}
\label{f.10}
\end{figure}

\begin{figure}
\begin{center}
\mbox{\epsfig{file=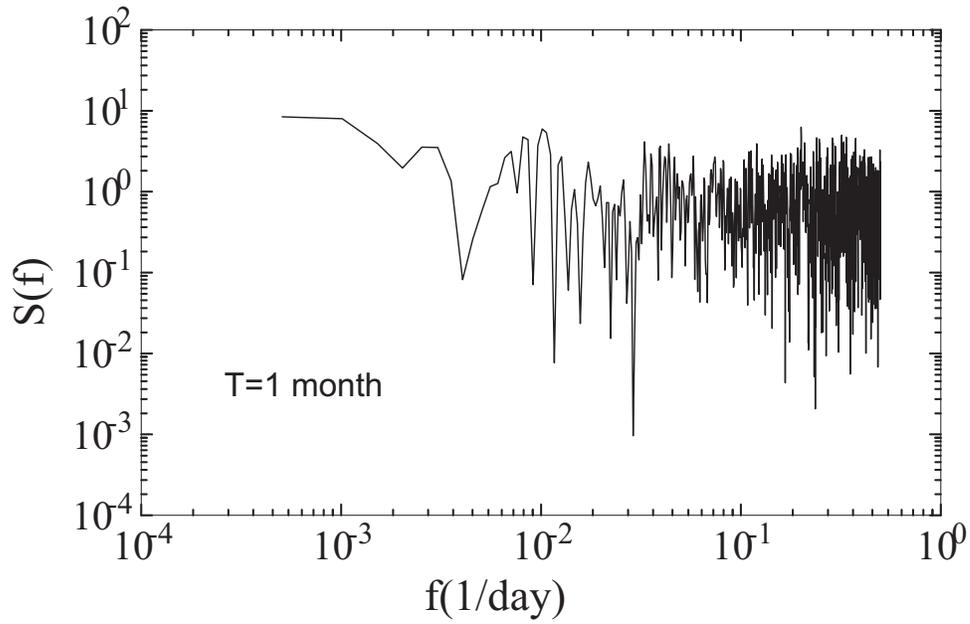,width=5in,angle=0}}
\end{center}
\caption{Power spectrum of $\Delta r(T,t)$ for the EURO with $T=1$
month and $\Delta t$=$1$ day.}
\label{f.11}
\end{figure}

\begin{figure}
\begin{center}
\mbox{\epsfig{file=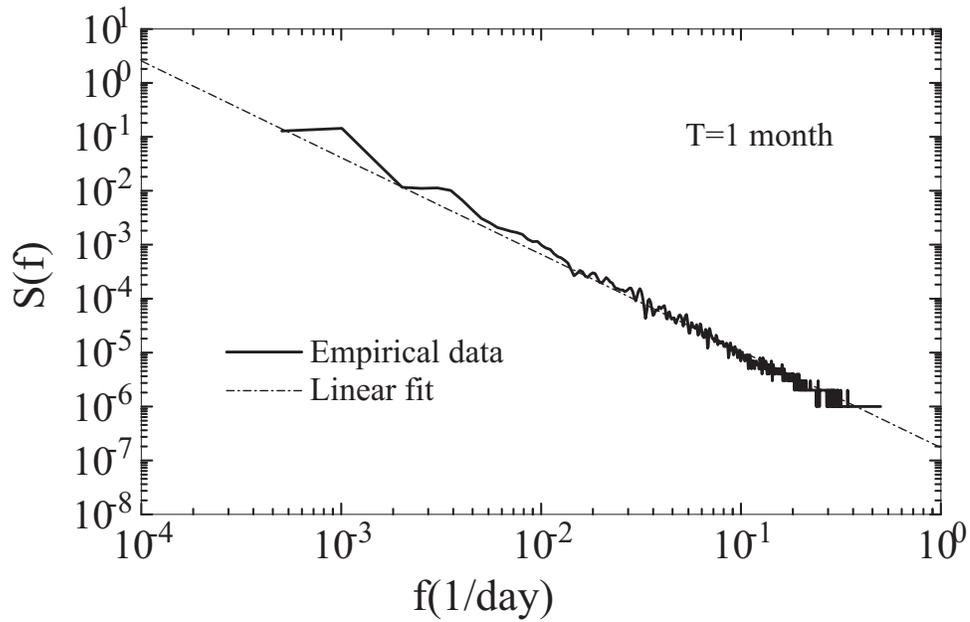,width=5in,angle=0}}
\end{center}
\caption{Power spectrum of $r(T,t)$ for the Sterling Pound with
$T=1$ month.}
\label{f.12}
\end{figure}

\begin{figure}
\begin{center}
\mbox{\epsfig{file=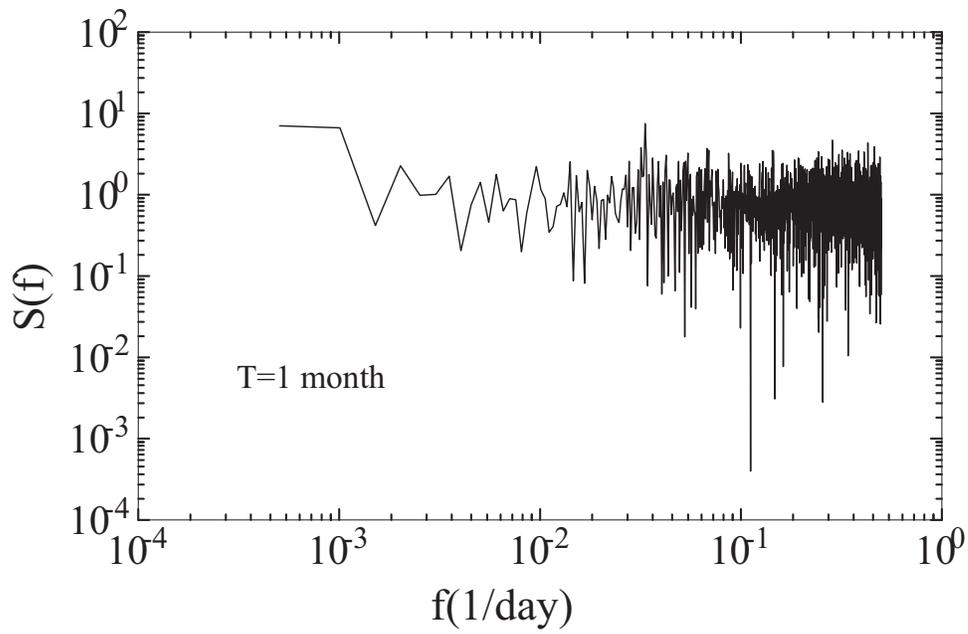,width=5in,angle=0}}
\end{center}
\caption{Power spectrum of $\Delta r(T,t)$ for the Sterling Pound
with $T=1$ month and $\Delta t$=1 day.}
\label{f.13}
\end{figure}

The power spectra, $S(f)$, for both $r(1,t)$ and $\Delta r(1,t)$
are reported in Figs.\ref{f.10} and \ref{f.11} for the EURO and in
Figs.\ref{f.12} and \ref{f.13} for the Sterling Pound. For $r$ the
spectral density shows a power law behavior. A linear fit gives a
slope value $\alpha={-1.80 \pm 0.02}$ and $\alpha={-1.79 \pm
0.01}$ for EURO and Sterling Pound, respectively. A similar result
holds for the other maturities. Therefore, we argue that the power
spectrum analysis for $r(T,t)$ indicates a stochastic process with
spectral components decreasing as $S(f) \sim {f}^{\alpha}$
(Feller (1971) \cite{Feller}). As for the increments, in Figs.\ref{f.11}
and
\ref{f.13}, the power spectrum is flat, typical of a white noise
process without seasonality peaks. These results are also corroborated by
a similar analysis
performed on Eurodollars interest rates for a longer time period (Di
Matteo (2001) \cite{DiMatteo}).

\section{Discussion} \label{s.4}

In the previous section, we have shown that the daily increment of
the interest rate series is non-Gaussian and follows a
leptokurtic distribution. The problem arises of how derivatives
written on interest rates can be evaluated, given that the usual
Gaussian white-noise assumption of many models is not satisfied.
Indeed, a first partial answer, is that the central limit theorem
ensures that, after a sufficiently long time, the increment
distribution will tend to a Gaussian distribution. However, if the
time horizon of derivative evaluation is not appropriate,
deviations from the Gaussian behavior may lead to a dramatic
underestimate of large increments with a consequent improper risk
coverage.

Although well studied in mainstream
finance, this problem has received much attention in recent times, within
the community of physicists working on financial problems
(Bouchaud (1997), Bouchaud (1994) \cite{BouchaudPot,Sornette 1994}). In
particular, Bouchaud and
Sornette (Bouchaud (1994) \cite{Sornette 1994}) suggested the direct use
of the
historical probability measure, rather than the equivalent
martingale measure for evaluating options. In this way, one gets
an option price depending on the expected rate of returns, a
consequence which is not fully desirable due to the subjective
character of that rate. Assessing trends is a difficult
task, as they depend on decision taken by Central Banks (as shown
in Figs.\ref{f.3} and \ref{f.4}) and are based on macroeconomic
effects. Thus, martingale methods could prove more
reliable. As early as 1977, some years after the seminal paper of
Black and Scholes, Parkinson generalized their approach to option
pricing and explicitly took into account leptokurtic distributions
(Parkinson (1977) \cite{Parkinson}). More recently, Boyarchenko and
Levendorskii have
studied the problem of option pricing in the presence of a
specific distribution which seems to fit well the empirical data
in many instances: the truncated L\'evy distribution
(Boyarchenko (2000), Koponen (1995) \cite{Boyarchenko,Koponen}). Along
these lines, we believe, it will
be possible to develop a consistent option pricing model taking
into account the leptokurtic character of the empirical short to
mid-term interest rate distributions.

\section*{Acknowledgements}

We thank the Risk Management division of Banca Intesa (Milano) for
providing us the data. T. Di Matteo wishes to thank Prof. Sandro
Pace and Tomaso Aste for fruitful discussions and support. This
work was partially supported by INFM (Istituto Nazionale per la
Fisica della Materia) FRA project - $AMF_2$: Application of
Physical Methods to Financial Market Analysis.

\newpage
\listoffigures
\newpage

\vspace*{1cm}

\vspace*{1cm}

\vspace*{1cm}

\vspace*{1cm}

\vspace*{1cm}

\vspace*{1cm}

\vspace*{1cm}

\vspace*{1cm}

\vspace*{1cm}

\vspace*{1cm}

\vspace*{1cm}

\vspace*{1cm}

\vspace*{1cm}

\end{document}